# Recombination and localization: unfolding the pathways behind conductivity losses in Cs$_2$AgBiBr$_6$ thin films

Huygen J. Jöbsis[1], Valentina M. Caselli[2], Sven H. C. Askes[3], Erik C. Garnett[3], Tom J. Savenije[2], Freddy T. Rabouw[1*], Eline M. Hutter[1*]

1. Inorganic Chemistry and Catalysis, Department of Chemistry, Utrecht University, Princetonlaan 8, 3584 CB Utrecht, the Netherlands
2. Department of Chemical Engineering, Delft University of Technology, van der Maasweg 9, 2629 HZ Delft, the Netherlands
3. Center for Nanophotonics, AMOLF, 1098 XG Amsterdam, the Netherlands

*f.t.rabouw@uu.nl

*E.M.Hutter@uu.nl

**Abstract**

Cs$_2$AgBiBr$_6$ (CABB) has been proposed as a promising non-toxic alternative to lead halide perovskites. However, low charge carrier collection efficiencies remain an obstacle for the incorporation of this material in optoelectronic applications. In this work, we study the optoelectronic properties of CABB thin films using steady state and transient absorption and reflectance spectroscopy. We find that optical measurements on such thin films are distorted as a consequence of multiple reflections within the film. Moreover, we discuss the pathways behind conductivity loss in these thin films, using a combination of microsecond transient absorption and time-resolved microwave conductivity spectroscopy. We demonstrate that a combined effect of carrier loss and localization results in the conductivity loss in CABB thin films. Moreover, we find that the charge carrier diffusion length and sample thickness are of the same order. This suggests that the material's surface is an important contributor to charge-carrier loss.

Lead-halide perovskites exhibit excellent light absorption and emission properties and micrometer-long charge carrier diffusion length, making them interesting for optoelectronic applications.[1–3] However, the presence of lead and iodide ions raises toxicity concerns.[4] In addition, many of the lead-based perovskites have poor stability in aqueous environments, limiting their application in *e.g.* photocatalysis.[5] Cesium silver bismuth bromide ($Cs_2AgBiBr_6$, CABB) has been suggested as a less toxic and more stable alternative for the remarkably performing lead-halide perovskites in optoelectronic application.[6–10]

Microsecond-long carrier lifetimes, increased stability in water and improved photostability with respect to high performing lead-containing analogues, have triggered an extensive research effort into CABB.[8,9,11–14] Its applicability in solar conversion or lighting applications is still limited due to the weak absorption and emission caused by the indirect bandgap.[15,16] On the other hand, the material performs well as an X-ray detector.[17,18] This can be understood from the presence of bismuth, causing efficient X-ray attenuation, and the long carrier lifetimes, which are desirable for charge extraction. More recently, CABB was also used for photocatalytic reactions such as light-driven $CO_2$ conversion and $H_2$ generation from hydrobromic acid.[5,19] However, in all of these applications, the charge extraction efficiency remains limited by the short diffusion length of the minority carrier. The limited extraction has been ascribed to the high trap density in CABB single crystals compared to lead-containing perovskite single crystals.[20] In a recent study, Wright *et al.* identified charge carrier self-localization to a small polaronic state with a localization rate of *ca.* 1 ps$^{-1}$ to be intrinsic to CABB.[21] Such fast localization rates would be detrimental to its application in optoelectronic devices. However, temperature-activated delocalization results in appreciable carrier mobilities at room temperature.[21] Moreover, mobile carriers were observed for microseconds after photoexcitation at elevated temperatures timescale,[20] highlighting CABB as a potential alternative for lead-halide analogues.

In this work, we study the charge-carrier dynamics in CABB thin films on nanosecond-to-microsecond timescales using a combination of transient absorption (TA) and time-resolved microwave conductivity (TRMC) spectroscopy. TA experiments present long-lived carriers ranging over several microseconds, while TRMC measurements on the same thin films show that all charge carriers are immobilized within 200 nanoseconds. In the first tens of nanoseconds the charge carrier mobility remains almost constant, indicating that the intensity drop of the TRMC trace on this timescale is the result of carrier loss, after which it drops due to localization of remaining carriers. So that we conclude that the conductivity loss in CABB thin films is the result of both carrier loss and localization. A comparison of the bleach recovery dynamics of the direct and indirect absorptions shows that photogenerated electrons are lost more rapidly than holes. Carrier diffusion lengths of the same order of magnitude as the sample thickness, suggest a dominant role of the material's surface in charge-carrier loss.

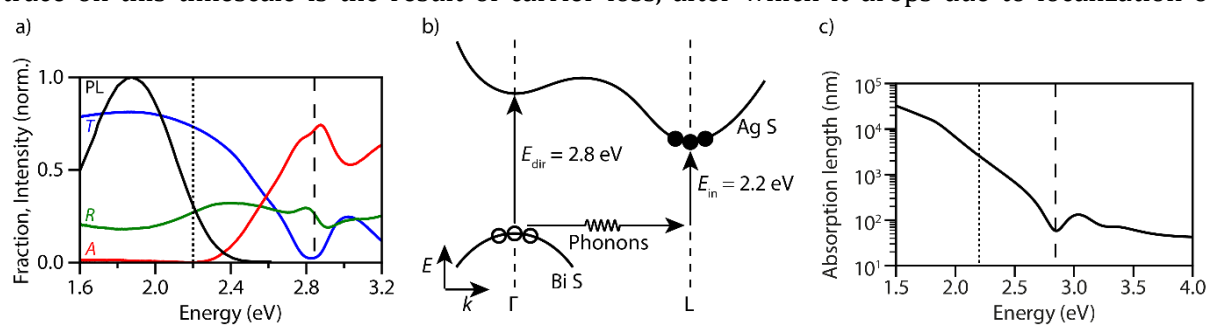

**Figure 1: Optical properties of CABB thin films.** a) The fraction of transmitted (*T*, blue line), reflected (*R*, green line) and absorbed (*A*, red line) light, as well as the photoluminescence intensity (PL, black line, $\lambda_{exc}$ = 440 nm) as a function of photon energy for a 100-nm-thick CABB film on glass. b) Schematic representation of the band structure of CABB, showing the direct and indirect absorption processes. c) The absorption length as a function of photon energy calculated based on ellipsometry data. The dotted and dashed line in panel a and c represent the indirect bandgap and excitonic transition in CABB.

We prepared CABB thin films of *ca.* 100 nm thick, following the method by Li *et al.*.[22] In short, a mixture of CsBr, AgBr and BiBr$_3$ in dimethyl sulfoxide was spin coated on an optical substrate, allowed to dry, and annealed. Experimental details and crystallographic data (Fig. S1) are provided in the Supporting Information. Atomic force microscopy shows a homogeneous film coverage with a root-mean-square roughness of 7.3 nm (Fig. S2).

The steady-state optical properties were determined using an UV-vis spectrometer with an integrating sphere (see experimental methods). The transmittance ($T$) spectrum, *i.e.* fraction of light transmitted as a function of photon energy, shows a characteristic dip around 2.8 eV (blue line in Fig. 1a). This energy is significantly larger than the energy of the bandgap of CABB, which is indirect and therefore does not cause a distinct dip in the transmittance spectrum. The feature at 2.8 eV is instead often attributed to the excitonic transition between conduction and valence band at the Γ point in the dispersion diagram (Fig. 1b).[23] We attribute the slow decrease in the transmittance, starting at 2.2 eV and increasing towards higher energies, to indirect absorption. Indeed, the indirect-bandgap energy of CABB single crystals,[11,24] thin films[8,9] and nanocrystals[25,26] is reported ranging from 1.95 to 2.3 eV. The photoluminescence (PL) (black line, Figure 1a) is considerably red-shifted compared to the indirect-bandgap absorption and relatively broad (centered at 1.9 eV, FWHM of 515 meV). The red-shift is known to be due to strong exciton phonon coupling.[8,11,21,27] Recent work suggests that the charge carrier recombination pathway in CABB proceeds via color centers.[21,24]

We observe oscillations in the reflectance ($R$; fraction of light reflected) spectrum (green line in Fig. 1a). These must be due to a combination of absorption resonances of CABB and the film's dielectric properties, which determine interference effects. The real part of a material's dielectric function peaks at frequencies just below an absorption resonance, while it shows a minimum at frequencies just above the resonance. As the reflectance of a bulk material scales with the dielectric contrast with air, it should peak just below a strong absorption resonance and dip at energies above the resonance. Indeed, $R$ of our CABB film shows such a wiggle feature around the excitonic resonance at 2.8 eV (Fig. 1a). A similar feature is not obvious around the indirect bandgap transition at 2.2 eV, which is likely the result of interference between multiple reflections within the film distorting the reflectance spectrum. Interference becomes an especially important factor determining the film's reflectivity in the spectral range below 2.8 eV, where the absorption of CABB is weak compared to the film thickness of 100 nm (Fig. 1c; calculated based on ellipsometry data Fig. S3) so multiple reflections of the light are possible.

Next, we study the change in optical properties using pump-probe TA spectroscopy experiments in transmission mode on a microsecond timescale (for a schematic of the TA setup see Fig. S4 in the Supporting Information).[28] The transient transmittance ($\Delta T/T$) shows a distinct bleach at 2.8 eV, which is present for excitation at energies above as well as below the direct transition (compare blue and green lines in Fig. 2a*).* The bleach must therefore be due to holes near the top of the valence band, which can be photoexcited with either direct or indirect excitation. For excitation at 355 nm an additional weak and broad bleach feature between 1.8 and 2.2 eV is present. This energy range comprises the reported values for the indirect bandgap of CABB. We propose that this signal is the result of a combination of absorption and reflection effects, which we will discuss below. The broad bleach signal is not as evident for excitation at 532 nm because of the low signal to noise ratio.

In reflection mode we study the change in reflectance of our CABB thin films, again for excitation at 355 and 532 nm (Fig. 2b). The $\Delta R/R$ spectrum displays much stronger features in the indirect-absorption region (<2.7 eV) than the $\Delta T/T$ spectrum, highlighting the potential to extract information on indirect-absorption transitions form reflectivity measurements. However, the interpretation of the spectra is not straightforward. For both excitation wavelengths we observe an inflection point in $\Delta R/R$ at 2.8 eV: decreased reflectance just below and increased reflectance just above the absorption resonance. This results in a minimum in $\Delta R/R$ at 2.75 eV followed by a maximum at 2.9 eV. This spectral shape is a consequence of the bleach of the absorption transition. Bleaching the absorption flattens the real part of the material's dielectric function around the absorption resonance and suppresses the wiggle in the reflectivity spectrum (see discussion above). For both excitation wavelengths a second inflection point is observed at lower energies. For excitation at 355 nm the inflection point occurs at 2.2 eV while it is at 2.4 eV for 532-nm excitation (compare blue and green arrows in Fig. 2b). While these are the approximate energies of the indirect bandgap in CABB, the different inflection points for different excitation wavelength indicate that inflections are not just due to a bleach of the indirect absorption. Instead, the $\Delta R/R$ spectrum at <2.7 eV is likely strongly distorted by interference effects in the thin film. Indeed, our CABB film is 100 nm thick and has a refractive index of approximately 2, so light of 1.6 eV picks up a phase of π upon back-and-forth reflection in the film, which increases to 1.7π at 2.7 eV, causing a slow variation of the interference over this range of photon energies. Considering the absorption length at the excitation wavelength (Fig. 1c), the 355-nm excitation will generate charge carriers mostly at the surface, while 532-nm excitation generates charge carriers more homogeneously. It is unclear to what extent this affects interference of probe light within the film. It is therefore not straightforward to assign the spectral features in the $\Delta R/R$ spectrum at <2.7 eV to any particular transition. We note that interference also affects a film's differential transmission $\Delta T$ (or its differential absorptance or differential absorbance) whenever the absorption length (Fig. 1c) is long compared to the film thickness, so care must be taken in the interpretation of spectra. Fig. S5 and S6 in the Supporting Information present the $\Delta T/T$ and $\Delta R/R$ spectra of our films as a function of delay time.

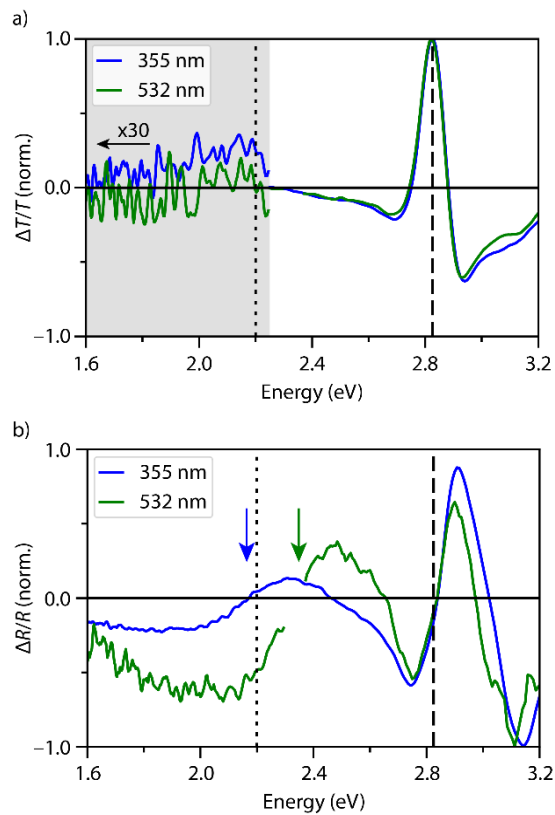

**Figure 2: Transient transmittance and reflectance spectral slices.** a) Normalized spectral slice of $\Delta T/T$ for excitation at 355 nm (blue) and 532 nm (green) after a delay time of 100 ns. In the grey-shaded area the data is magnified by a factor of 30. b) Normalized spectral slice of $\Delta R/R$ for excitation at 355 nm (blue) and 532 nm (green) after a delay time of 100 ns. The blue and green arrow indicate the inflection point of $\Delta R/R$ for excitation at 355 and 532 nm, respectively. In both panels the dotted and dashed lines respectively represent the indirect bandgap and excitonic transition of CABB.

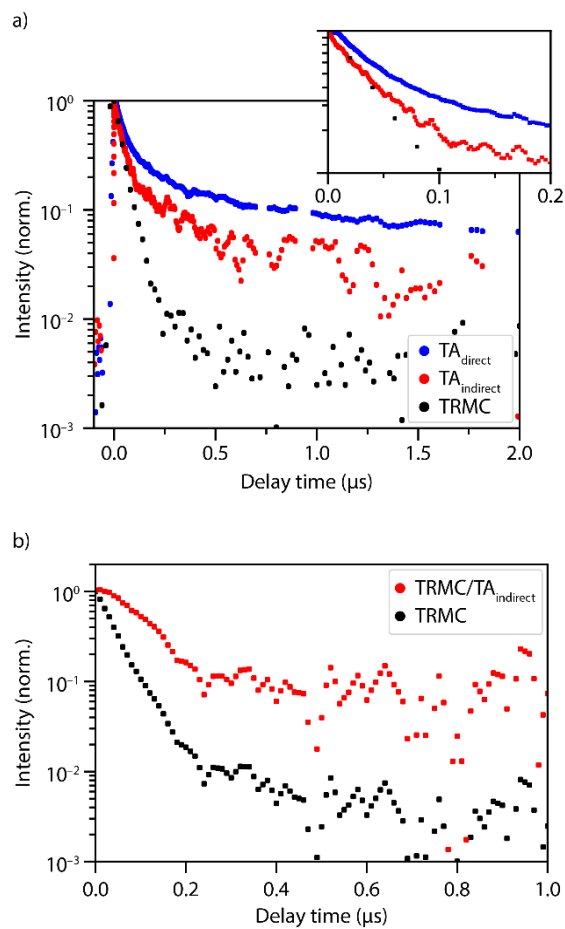

**Figure 3: TA and TRMC time traces.** a) Normalized TA time traces at 2.2 eV (red) and 2.8 eV (blue) ($\lambda_{exc}$ = 355 nm). Normalized TRMC time trace (black, $\lambda_{exc}$ = 355 nm). The inset is a zoom-in of the kinetics between 0 to 200 ns. b) The calculated mobility trace (red) by taking the ratio between the TRMC trace and the TA trace observed for the indirect transition. For clarity the TRMC trace in panel (a) is reproduced in black All traces are normalized to unity at $t$ = 0.

Finally, we study the charge carrier decay dynamics for our CABB thin films. The $\Delta T/T$ traces for both bleach signals (direct transition at 2.82 eV (blue) and indirect transition at 2.2 eV (red)) are shown for excitation at 355 nm in Fig. 3a. We observe a long-lived tail ranging over several microseconds in both measurements. Such slow decay dynamics in TA experiments on CABB thin films have previously been observed by Hoye *et al.*[11] In contrast, TRMC measurements (black in Fig. 3a) on the same CABB thin films show that 99% of free-carrier mobility is lost within 200 ns after photoexcitation. These observations indicate that although the photoexcited electrons and holes localize on timescales of tens of nanoseconds (losing mobility), recombination extends over several microseconds. The bleach of the indirect absorption drops considerably faster in the first 100 ns than that of the direct absorption. Typically, on this timescale the charge carriers have relaxed to the band edges so that the charge carriers occupy their corresponding bands as depicted in Fig. 1b. From this it follows that the bleach recovery dynamics of the indirect transition scales with both the electron and hole population, whereas the direct-absorption bleach depends only on the hole population. The decay dynamics do not depend significantly on excitation wavelength (Fig. S7).

The decay kinetics are fluence-independent for the TA and TRMC experiments (Fig. S8), which allows comparison of the observed traces. The TRMC signal probes the product of charge carrier concentration and mobility, whereas the TA signal scales only with the concentration. Assuming that all charge carriers contribute equally to the TA signal, independent of the degree of localization, the ratio between the TRMC and TA traces reveals the decay of overall mobility of photoexcited charge carriers. As discussed above the TA trace observed for the indirect transition scales with both the electron and hole population, so that comparison of this trace with the TRMC traces provides insight into the mobility over time. For delay times <40 ns the TRMC and TA trace observed for the indirect transition show similar decay kinetics (inset of Fig. 3a). This means that the mobility remains close to constant and, on this timescale, both carriers are mobile and the conductivity loss is the result of charge carrier loss. Interestingly, we observe a short-lived plateau in the bleach recovery dynamics of the direct transition (depending only on hole population), indicating that in the first 10 ns after photoexcitation the hole population remains close to constant. This suggests that the initial drop in TRMC signal over the first 10 ns is the result of a change in the electron population density. On the timescales from 10 to 40 ns, both electrons and holes are lost, which the mobility of remaining carriers is unaffected (Fig. 3b). Beyond 40 ns, the charge carriers localize and their mobility goes down. More experiments would be necessary to distinguish how much electrons and holes contribute to the overall conductivity.

With the use of the carrier mobility and the free carrier half lifetime, i.e. the time after which the TRMC signal has decayed by a factor two, we can estimate the carrier diffusion length.[29] Recent work on CABB report charge carrier mobilities in the range of 0.3 to 11 cm$^2$ V$^{-1}$ s$^{-1}$ at room teperature.[21,30] With a free carrier half lifetime of 35 ns we approximate the corresponding carrier diffusion length to be at least 160 nm (Table S1), which is on the same order as the sample thickness. In literature it has been suggested that surface related charge carrier recombination limits the photoconductance in CABB thin films.[20] Based on our observation discussed above we can similarly conclude that surface trapping leads to a decay in photoconductivity. This would mean that the photoconductance can be increased through, *e.g.* surface passivation, further improving the potential of CABB-based optoelectronic application.

In this work we studied the steady-state and transient optical properties of CABB thin films. The steady-state transmission spectrum on the thin films clearly shows the direct absorption transition at 2.8 eV. The indirect bandgap is, however, not distinctly present. The reflectance $R$ and transient reflectance $\Delta R/R$ spectra show features in the indirect-bandgap region. They are likely dominated by multiple interference effects in the thin CABB film and difficult to assign to any particular electronic transition. Further research into the influence of such interference effects on $R$ and $\Delta R/R$ is required, including possible influences of a nonhomogeneous charge carrier distribution, as this will ultimately affect the absorption for thin films in spectroscopic studies as well as applications.

TA spectroscopy revealed that the charge carrier density decays on a 40-nanosecond timescale but a fraction of the photogenerated holes near the valence band maximum have a lifetime ranging over several microseconds. TRMC measurements showed that these long-lived carriers are, however, not mobile. Comparison of the TRMC and TA traces shows that the conductivity loss is the result of a combined effect of charge carrier loss and localization. Finally, we find that the charge carrier diffusion length is of the same order of magnitude as the film thickness, suggesting that surface trapping lead to a decay in photoconductivity. This means that the photoconductance can be increased through improved material design, further enhancing the potential of CABB for optoelectronic applications.

**X-ray diffraction crystallography.** X-ray diffraction (XRD) patterns were obtained with a Bruker-AXs D8 Phaser powder X-ray diffractometer in Bragg-Brentano geometry, using Cu K$\alpha_{1,2}$ = 1.54184 Å, operated at 40 kV. The measurements used a step size of 0.01° and a scan speed of 1 s, with a 2 nm slit of the source.

**Steady-state optical properties.** The transmittance, absorptance and reflectance were determined using a Perkin-Elmer lambda UV-Vis-NIR_lambda950S with integrating sphere. The transmittance (fraction of photons transmitted, $T$) was determined by placing the CABB thin film at the entrance of the integration sphere. The sum of reflectance (fraction of photons reflected; $R$) and transmittance was determined by placing the thin film inside the sphere under an angle of 15°. Now the reflectivity and absorptivity (fraction of absorbed light, $A$) were calculated using

$$R = (R + T) - T$$

and

$$A = 1 - (R + T)$$

The photoluminescence spectrum has been recorded with a FLAME-S-VIS-NIR Ocean Optics spectrometer upon photoexcitation by CPS405 ThorLabs laser diode at 405 nm, in a home build set-up.

**Determination of optical constants.** The optical constants of the CABB layer were extracted using ellipsometry on a J. A. Woollam Variable Angle Spectrometric Ellipsometer (VASE) between 300 and 1700 nm with 2 nm wavelength spacing and at 5 different angles around the Brewster angle of ~65° (at 55°, 60°, 65°, 70°, and 75° angles). The layer thickness was measured using a KLA-Tencor alpha-step 500 surface profiler. The surface roughness (7.3 nm root mean square) was measured using Atomic Force Microscopy (AFM), performed on a Veeco Dimension 3100 AFM, and calculated using Gwyddion software. From these combined data, the optical constants of the CABB layer were extracted using CompleteEASE software by employing a generic oscillator model that included two Tauc-Lorentz oscillators to model the direct and indirect bandgaps, a Lorentz oscillator for the exciton peak at 2.8 eV, and a Gaussian oscillator for the peak at 3.2 eV.

**Transient absorption spectroscopy.** The transient absorption measurements were carried out using an EOS multichannel pump probe transient absorption spectrometer (Ultrafast Systems LLC.). The second (532 nm) or third harmonic (355 nm) of a Nd:YAG laser (1064 nm, 650 ps FWHM pulse duration, 1 kHz) was used as the excitation source. Laser pulses with an energy of 4.5 uJ or 5.6 uJ were focused to a spot size of $3 \times 10^{-4}$ cm$^{-2}$, corresponding to a fluence of $2.7 \times 10^{16}$ (at 355 nm) or $5.0 \times 10^{16}$ (at 532 nm) photons cm$^{-2}$ per pulse.

The probe pulse was generated using a LEUKOS super continuum light source (200-2400 nm, 200 mW, < 1ns) operating at 2 kHz. The white light is split into a probe and a reference beam, which is used to correct for fluctuations in the probe intensity. The reference and probe beam transmission spectra are both detected using fiber optics coupled multichannel spectrometer with a CMOS sensor (spectral resolution of 1.5 nm).

**Time resolved microwave conductivity spectroscopy.** Time resolved microwave conductivity measurements have been carried out upon photoexcitation at 355 nm. The excitation is performed by a Nd:YAG laser pulse of 3 ns, at 10 Hz repetition rate. The light intensity has been varied in the range $10^{11}$ to $11^{13}$ photons/cm$^2$ per pulse. Monochromatic microwaves, in the range 8.2 to 12.2 GHz, are generated by a voltage-controlled oscillator, and the measurements are performed at the resonance frequency of approximately 8.5 GHz. The photoconductance signal ($\Delta G$) is obtained from the normalized change in microwave power ($P$) due to interaction with free charge carriers, as:

$$\frac{\Delta P(t)}{P} = -K \Delta G(t)$$

where $K$ is a pre-determined sensitivity factor.

## Acknowledgement

H.J.J. and E.M.H. acknowledge funding from the Dutch Research Council (NWO) under the grant number VI.Veni.192.034. H.J.J. and E.M.H. are further supported by the Advanced Research Center Chemical Building Blocks Consortium (ARC CBBC). V.M.C. and T.J.S. acknowledge funding from the Dutch Research Council (NWO), grant number 739.017.004. S.H.C.A. acknowledges funding from the Dutch Research Council (NWO), grant number VI.Veni.192.062.

Recombination, Diffusion, and Radiative Efficiencies. *Acc. Chem. Res.* **49**, 146–154 (2016).

**Supporting information**

# Recombination and localization: unfolding the pathways behind conductivity losses in Cs$_2$AgBiBr$_6$ thin films

Huygen J. Jöbsis[1], Valentina M. Caselli[2], Sven H. C. Askes[3], Erik C. Garnett[3], Tom J. Savenije[2], Freddy T. Rabouw[1*], Eline M. Hutter[1*]

1. Inorganic Chemistry and Catalysis, Department of Chemistry, Utrecht University, Princetonlaan 8, 3584 CB Utrecht, the Netherlands

2. Department of Chemical Engineering, Delft University of Technology, van der Maasweg 9, 2629 HZ Delft, the Netherlands

3. Center for Nanophotonics, AMOLF, 1098 XG Amsterdam, the Netherlands

**Corresponding Authors**

*F.T.Rabouw@uu.nl, *E.M.Hutter@uu.nl

Experimental Information

**Chemicals.** 99.9% silver bromide (AgBr, Strem)**,** 99.999% cesium bromide (CsBr, Sigma-Aldrich), 99.998% bismuth bromide ($BiBr_3$, Sigma-Aldrich), 99.9% anhydrous dimethyl sulfoxide (DMSO, Sigma-Aldrich).

**CABB thin film synthesis.** In a nitrogen filled glovebox 1.5 mmol CsBr, 0.75 mmol AgBr and 0.75 mmol $BiBr_3$ were dissolved in 1.5 mL DMSO by heating the mixture to 70 °C for 1 h until a clear yellow solution was obtained. 15 mm x 15 mm borosilicate glass optical substrates were cleaned by 5 min $O_2$ plasma cleaning. 80 μL perovskite solution was spin coated on the optical substrate by spinning at 4000 rpm for 40 seconds. The sample was allowed to rest for 15 minutes at room temperature before annealing at 250 °C for 5 minutes.

Figure S1 - Crystallographic data
X-Ray Diffraction patterns of the $Cs_2AgBiBr_6$ thin film. The reference diffractogram was
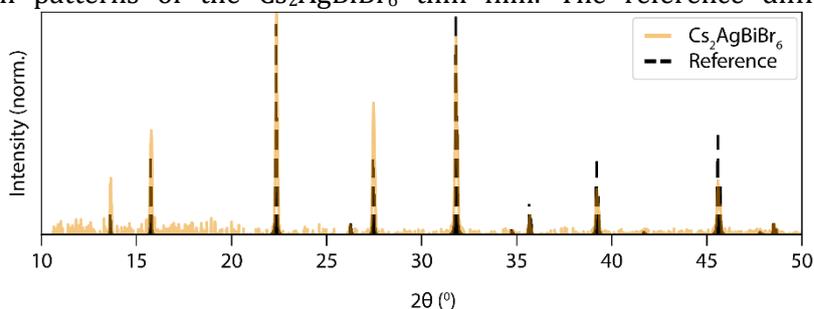
reproduced from ref[1].

Figure S2 –Atomic Force Microscopy
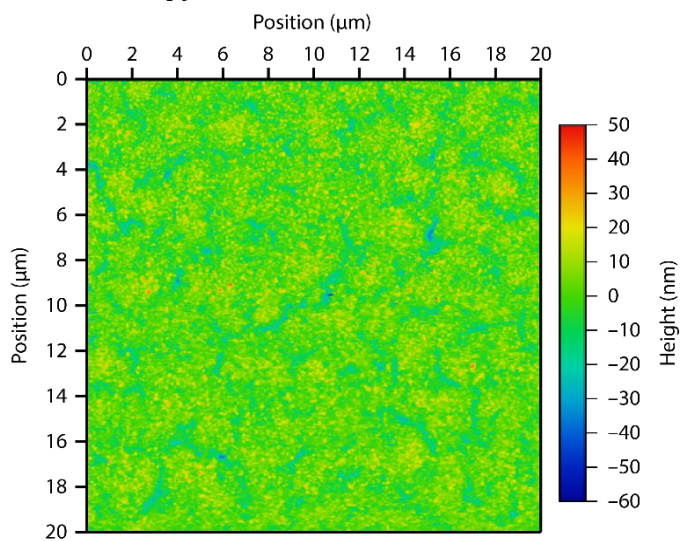

Atomic force microscopy image of an area of 400 μm² of the CABB thin film, showing a homogeneous films coverage with a root-mean-square roughness of 7.3 nm.

Figure S3 – Ellipsometry spectroscopy

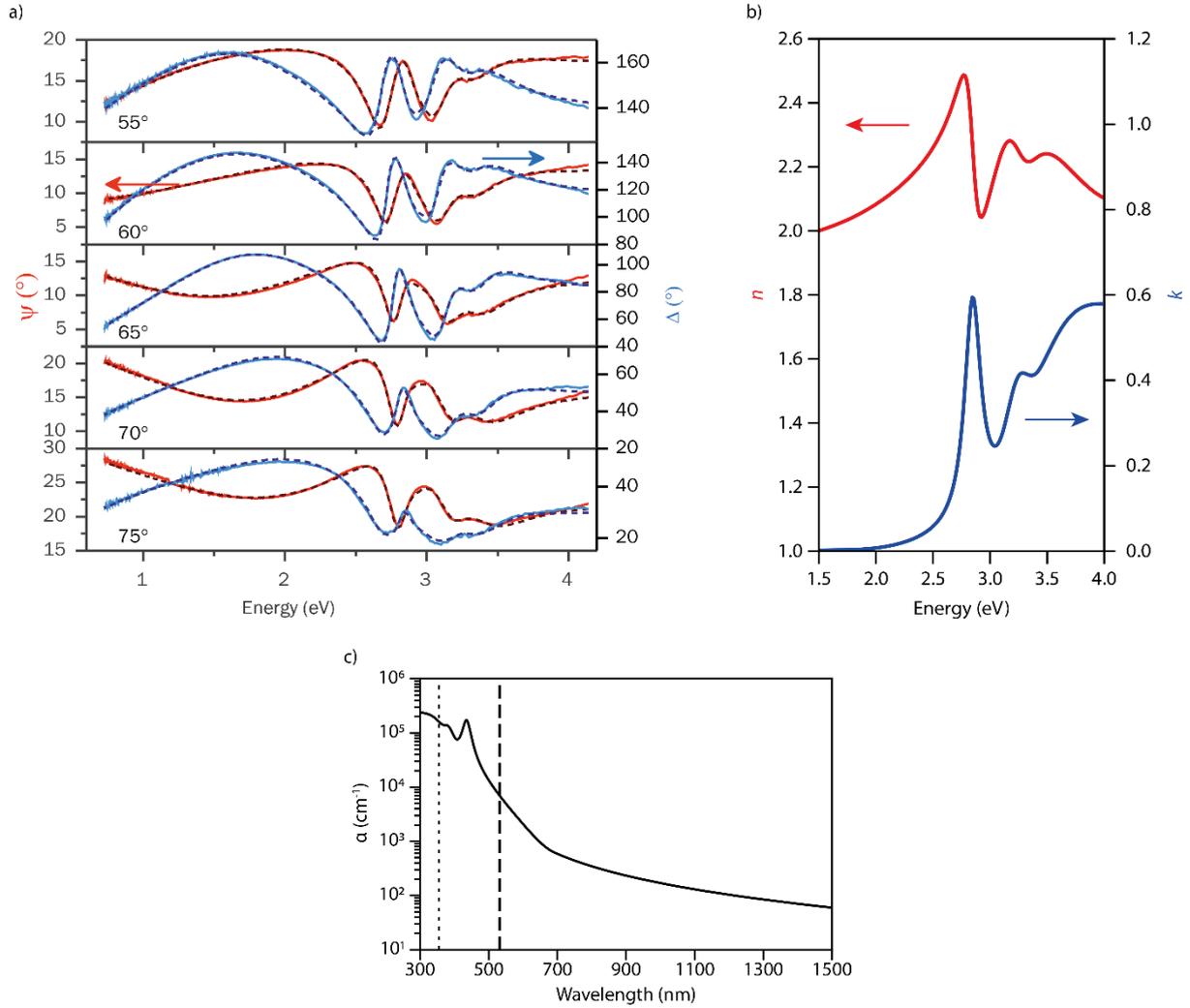

a) Complex reflection (amplitude (red, Ψ) and phase difference (blue, Δ) measured at 5 different angles around Brewster angle with 2 nm resolution. The optical constants of the CABB layer were extracted using CompleteEASE software by employing a generic oscillator model that included two Tauc-Lorentz oscillators to model the direct and indirect bandgaps, a Lorentz oscillator for the exciton peak at 2.8 eV, and a Gaussian oscillator for the peak at 3.2 eV (dashed black line). b) The refractive index (red, $n$) and extinction coefficient (blue, $k$) extracted from ellipsometry experiments. c) With the use of $k$ the absorption coefficient ($\alpha$) was calculated using

$$\alpha(\lambda) = \frac{4\pi k}{\lambda} \tag{1}$$

The dashed and dotted lines represent the excitation wavelengths at 355 and 532 nm, respectively.

Figure S4 – Setup of transient absorption experiments in transmission and reflection mode

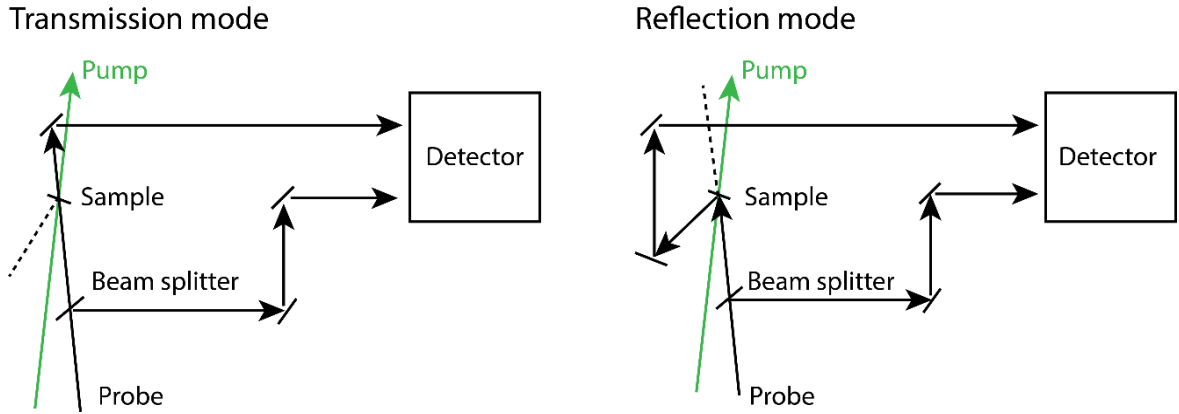

The probe pulse is split into two paths by a beam splitter. One part passes through the sample before entering a detection fiber, while the other is directed to a second fiber directly. The transmitted intensity is corrected for fluctuations in the probe intensity to obtain the differential transmittance $\Delta T/T$ of the sample:

$$\frac{\Delta T}{T} = ln\left(\frac{I^*}{I^*_{ref}} \frac{I_{ref}}{I}\right) \qquad (2)$$

where $I$ is the intensity of light transmitted through the sample without pump pulse and $I_{ref}$ the corresponding intensity in the reference channel, while $I^*$ and $I^*_{ref}$ are the intensities after pumping the sample. Similarly, we calculate the differential reflectance $\Delta R/R$ from a transient reflection measurement using

$$\frac{\Delta R}{R} = ln\left(\frac{I^*}{I^*_{ref}} \frac{I_{ref}}{I}\right) \qquad (3)$$

where $I$ and $I^*$ are the intensities of reflected light without and with pump pulse, respectively.

Figure S5 – Transient transmittance (Δ$T/T$)

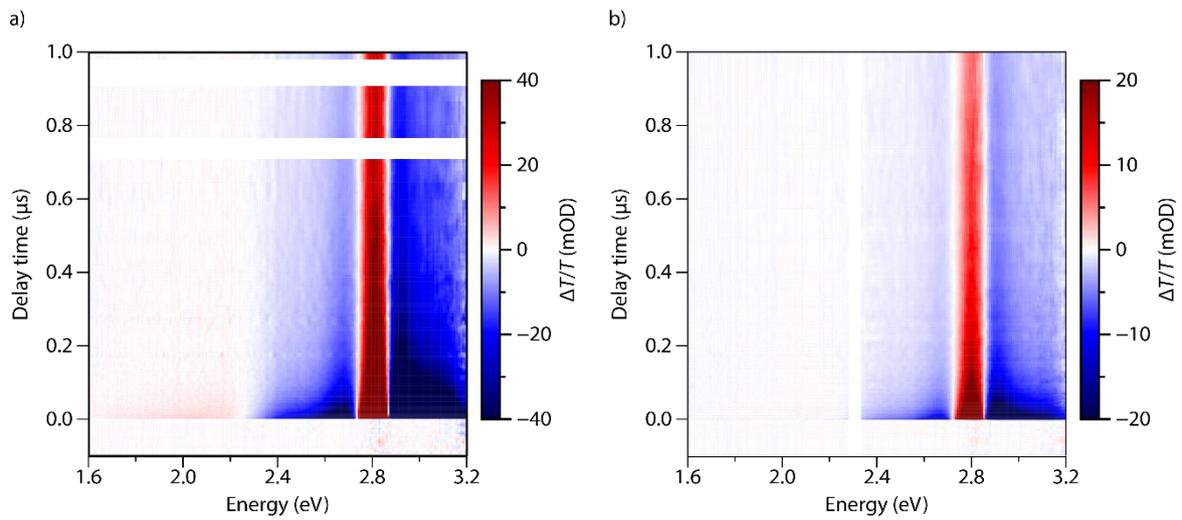

a) TA spectra as a function of delay time and photon energy for excitation at 355 nm (estimated charge carrier density $n$ = 0.209 nm$^{-3}$ per pulse). b) TA spectra as a function of delay time and photon energy for excitation at 532 nm (estimated charge carrier density $n$ = 0.202 nm$^{-3}$ per pulse). The gap in the data at 2.3 eV is due to excitation source.

Figure S6 – Transient reflectance (Δ$R/R$)

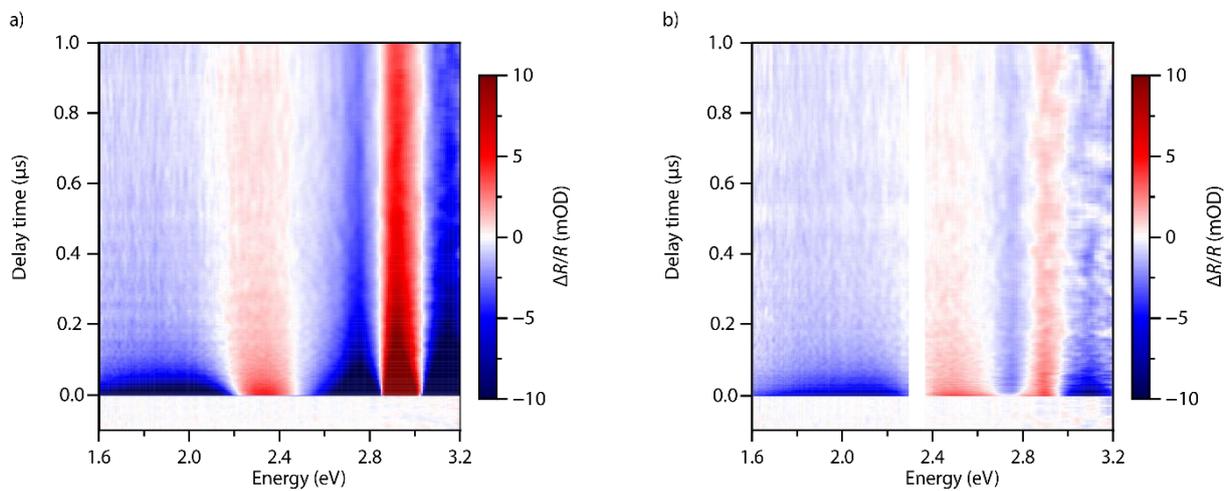

a) TR spectra as a function of delay time and photon energy for excitation at 355 nm ($n$ = 0.209 nm$^{-3}$ per pulse). b) TR spectra as a function of delay time and photon energy for excitation at 532 nm ($n$ = 0.202 nm$^{-3}$ per pulse). The gap in the data at 2.3 eV is due to excitation source.

Figure S7 – TA trace of direct absorption transition

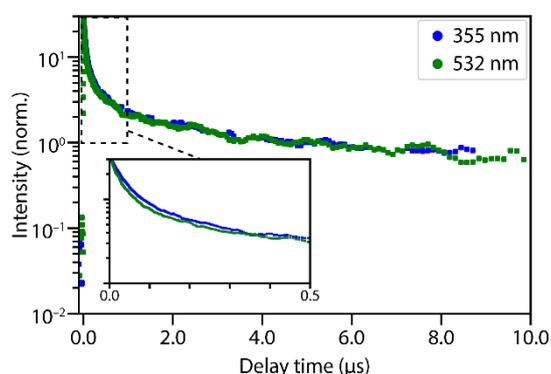

The bleach recovery dynamics of the direct absorption transition for excitation at 355 and 532 nm. The traces are normalized at 5 µs, showing similar decay dynamics. The inset shows a zoom in for the decay dyanmics on short timescales.

Figure S8 – Fluence-independent TA and TRMC traces

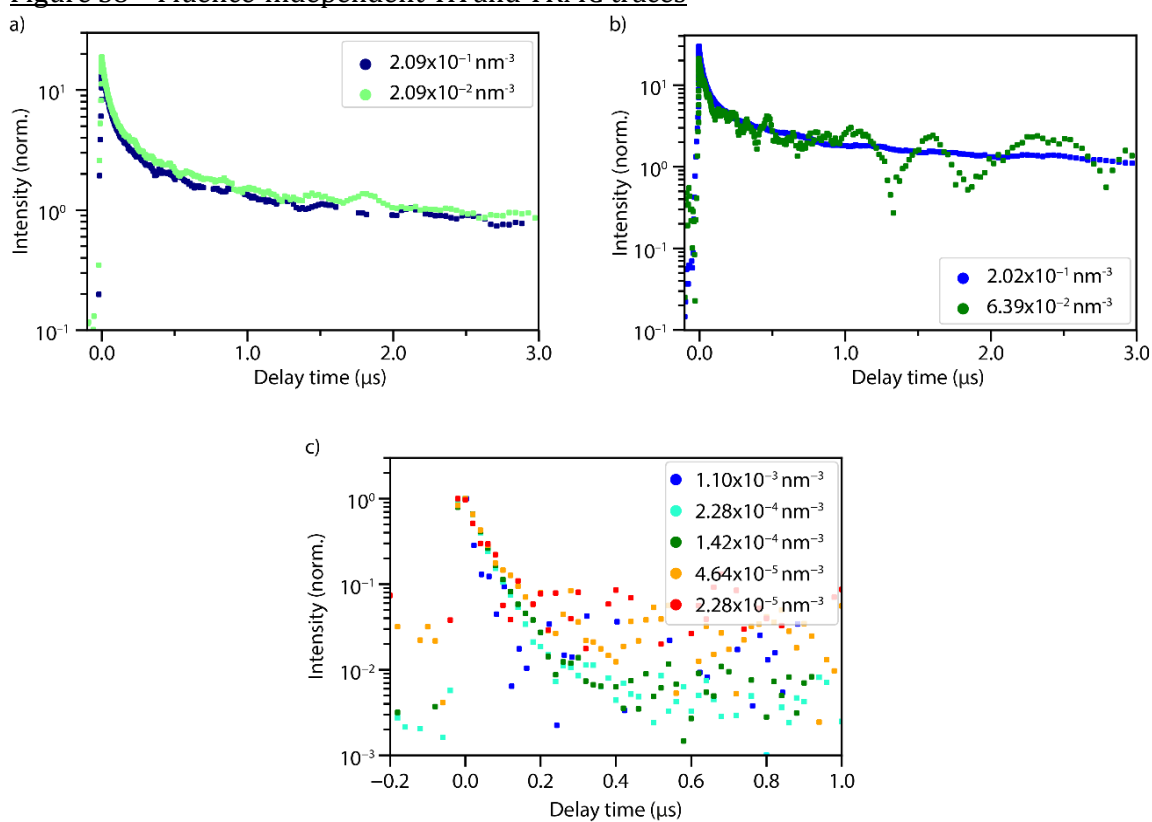

The bleach recovery dynamics at 2.8 eV for excitation at (a) 355 nm and (b) 532 nm for different excitaion densities. The time traces are normalized at 2 µs, showing that the decay kinetics are nearly independent of excitation density. The excitation density is given as a number of excitation per cubic nanometer per pulse. Panel c shows the normalized TRMC signal for different excitation densities.

Table S1 – Diffusion length

The diffusion length was estimated from the half lifetime ($\tau_{1/2}$), *i.e.* the time after which the TRMC signal has dropped by a factor 2, and the mobility (reproduced from literature), using [2-5]

$$L_\text{D} = \sqrt{\mu \frac{k_B T}{e} \tau_{1/2}} \quad (4)$$

with $\mu$ the charge carrier mobility, $k_B$ the Boltzmann's constant, $T$ the temperature, $e$ the elementary charge.

| Sample | Half lifetime (ns) | Mobility (cm$^2$/(V s)) | $L_\text{D}$ (nm) |
|---|---|---|---|
| Cs$_2$AgBiBr$_6$ | 35 | 0.3–11 | 164–997 |